%% file: main.tex
\documentclass[envcountsame]{llncs}

\usepackage{bussproofs} 
\usepackage{wrapfig}
\usepackage{times}
\usepackage{proof}
\usepackage{amssymb}
\usepackage{amsmath}
\usepackage{url}
\usepackage{color}
\usepackage{multicol}
\usepackage{wasysym}
\usepackage[inline]{enumitem}
\setlist{nolistsep}

\usepackage{graphicx}
\usepackage{tikz}
\usetikzlibrary{calc}
\usepackage{algorithm}
\usepackage[noend]{algpseudocode}
\usepackage{eufrak}

\usepackage{siunitx}

\spnewtheorem{Definition}[definition]{Definition}{\bfseries}{}
\spnewtheorem{Theorem}[theorem]{Theorem}{\bfseries}{}
\spnewtheorem{Lemma}[theorem]{Lemma}{\bfseries}{}
\spnewtheorem{Corollary}[theorem]{Corollary}{\bfseries}{}
\spnewtheorem{Example}[definition]{Example}{\bfseries}{}     
\spnewtheorem{Algorithm_Custom}[algorithm]{Algorithm}{\bfseries}{}     
\spnewtheorem{Remark}[definition]{Remark}{\bfseries}{}  
\spnewtheorem*{Proof}{Proof}{\bfseries}{}
  
\makeatletter
\renewcommand\paragraph{\@startsection{paragraph}{4}{\z@}%
                       {-2\p@ \@plus -4\p@ \@minus -4\p@}%
                       {-0.5em \@plus -0.22em \@minus -0.1em}%
                       {\normalfont\normalsize\bfseries}}
\renewcommand\section{\@startsection{section}{1}{\z@}%
                       {-8\p@ \@plus -4\p@ \@minus -4\p@}%
                       {8\p@ \@plus 4\p@ \@minus 4\p@}%
                       {\normalfont\large\bfseries\boldmath
                        \rightskip=\z@ \@plus 8em\pretolerance=10000 }}
\renewcommand\subsection{\@startsection{subsection}{2}{\z@}%
                       {-8\p@ \@plus -4\p@ \@minus -4\p@}%
                       {4\p@ \@plus 2\p@ \@minus 2\p@}%
                       {\normalfont\normalsize\bfseries}}                        

\makeatother

\newcommand{\partA}{\mathcal{A}}
\newcommand{\partB}{\mathcal{B}}
\DeclareMathOperator{\Prem}{Prem}
\DeclareMathOperator{\In}{In}
\DeclareMathOperator{\Out}{Out}
\DeclareMathOperator{\Dep}{Dep}
\DeclareMathOperator{\Sub}{Sub}
\newcommand{\lang}[1]{\mathcal{N}(#1)}
\newcommand{\infc}{r}
\newcommand{\infsys}{\mathfrak{S}}

\newcommand{\Vampire}{{\sc Vampire}}

\newcommand{\simpleAlg}{Simple-splitting-formula}
\newcommand{\linearAlg}{Linear-splitting-formula}
\newcommand{\splitByHeur}{Top-down-weighted-sum-heuristic}
\newcommand{\splitOptimally}{Weighted-sum-optimal}

\newcommand{\leaveout}[1]{}

\newcommand{\overlineshort}[1]{\mkern 1.0mu\overline{\mkern-1.0mu#1\mkern-1.0mu}\mkern 1.0mu}

\newenvironment{guide}{}{}

\title{Splitting Proofs for Interpolation\thanks{%
This work was supported by the ERC Starting Grant 2014 SYMCAR 639270,
the Wallenberg Academy Fellowship 2014 TheProSE,  the Swedish VR grant
GenPro D0497701 and the FWF projects S11403-N23 and S11409-N23.
We also acknowledge support from the FWF project W1255-N23. 
}}
\author{
 Bernhard Gleiss\inst{1} \and 
 Laura Kov{\'a}cs\inst{1,2} \and 
 Martin Suda\inst{1}
}
\institute{
TU Wien, Austria\and Chalmers University of Technology, Sweden
}

\begin{document}
\maketitle

\begin{abstract}


We study interpolant extraction from local first-order refutations.
We present a new theoretical perspective on interpolation
based on clearly separating the condition on logical strength of the formula
from the requirement on the common signature. This allows us to highlight 
the space of all interpolants that can be extracted from a refutation
as a space of simple choices on how to split the refutation into two parts.
We use this new insight to develop an algorithm for extracting interpolants
which are linear in the size of the input refutation and can be further
optimized using metrics such as number of non-logical symbols or quantifiers.
We implemented the new algorithm in first-order theorem prover \Vampire{}
and evaluated it on a large number of examples coming from the
first-order proving community. Our experiments give practical evidence
that our work improves the state-of-the-art in first-order interpolation.



%
%
%
%
%
%

\end{abstract}

\input{introduction}
\input{prelim}

\input{local_proof_interpolation}

\input{splitting_functions}
\input{related}

\input{experiments}
\input{conclusion}

\bibliography{bib}
\bibliographystyle{cs-abbrv++}

\end{document}

%% file: introduction.tex
\section{Introduction}
\label{sec:introduction}

Starting with the pioneering work of McMillan~\cite{McMillan03}, 
interpolation became a powerful approach in verification 
thanks to its use in predicate abstraction 
and model checking~\cite{DBLP:conf/cav/McMillan06,DBLP:conf/lpar/AlbertiBGRS12,Wies16}. 
To prove program properties over a combination of data structures, such as integers, arrays and pointers, 
several approaches based on theory-specific reasoning have been proposed, see e.g.~\cite{LahiriMehra05,CimattiGS08,Christ16}. 
While powerful, these techniques are limited to quantifier-free fragments of first-order logic. 
Addressing reasoning in full first-order theories, quantified interpolants are computed in~\cite{McMillan08,DBLP:conf/cade/KovacsV09,DBLP:conf/dagstuhl/ChristH10,Totla:2013} and further optimized with respect to various measures in~\cite{HoderKV12}.

In this paper, we address interpolation in full first-order logic and introduce a novel approach to generate interpolants, possibly with quantifiers. 
Our approach improves and simplifies the aforementioned techniques, in particular~\cite{DBLP:conf/cade/KovacsV09,HoderKV12}. 
In~\cite{DBLP:conf/cade/KovacsV09,HoderKV12}, the size of computed interpolants is in the worst case quadratic in the size of the proof and the generated interpolants may contain redundant subformulas. Our work addresses these issues and infers interpolants that are linear in the size of the proof and are much simpler than in~\cite{DBLP:conf/cade/KovacsV09,HoderKV12}. 
%
We proceed as follows. 
We separate the requirements on a formula being  an interpolant into a part restricting the logical strength of an interpolant  and a part restricting which symbols are allowed to be used in an interpolant. 
This way, we first handle formulas, called intermediants,  satisfying the requirements on the logical strength of interpolants, 
and only then we restrict the generated space of intermediants to the ones that satisfy the restriction on the interpolants signature. 

The work of~\cite{DBLP:conf/cade/KovacsV09} relies on so-called local proofs (or split proofs) and constructs interpolants by splitting local proofs into (maximal) subproofs. Splitting proofs is determined by the signature of formulas used in the proofs. We observed, however, that there are many ways to split a proof, resulting in interpolants that are different in size and strength. 
We therefore propose a general framework for splitting proofs
and using the boundaries of the resulting sub-proofs to construct the intermediants.
The key feature of our work is that the interpolants inferred from our various proof splits are linear in the size of the proof. When constructing interpolants from proof splits, we note that local proofs are exactly the ones that ensure that proof splits yield intermediants that satisfy 
the requirements of interpolants. Using local proofs and proof splits, we then describe a powerful heuristic and an optimality criterion how to choose the ``best''  proof split, and hence the resulting interpolant. 

\smallskip

\noindent{\bf Contributions.} The main contributions of this paper are as follows. 
\begin{itemize}
	\item We present a new algorithm for first-order-interpolation using local proofs in arbitrary sound inference systems. That is, our work can be used in any sound calculus and derives interpolants, possibly with quantifiers, in arbitrary first-order theories. 
	\item Our interpolation algorithm is the first algorithm ensuring that the size of the interpolant is linear in the size of the proof
	 while working with an arbitrary sound logical calculus. This result improves~\cite{DBLP:conf/cade/KovacsV09} and generalises the work of~\cite{McMillan08} to any sound inference system.

	\item We implemented our work in the \Vampire{} theorem prover~\cite{KovacsVoronkov:CAV:Vampire:2013} and evaluated our method on a large number of examples coming from the TPTP library~\cite{tptp}. Our experimental results confirm that our work improves the state-of-the-art in first-order interpolation. 
\end{itemize}
\noindent
The rest of this paper is structured as follows. 
The background notation on proofs and interpolation is covered in Section~\ref{sec:prelim}. 
We then show how to construct linear sized interpolants in Section~\ref{section:local_proof_interpolation}
and present optimisations to the procedure in Section~\ref{section:splitting_functions}. 
We compare to related work in Section~\ref{sec:related},
describe our experimental results in Section~\ref{sec:experiments}, 
and conclude in Section~\ref{sec:conclusion}.

%% file: prelim.tex
\section{Preliminaries}
\label{sec:prelim}
This section introduces the relevant theoretical notions to our work. 

\paragraph{\bf Formulas.}
We deal with standard first-order predicate logic with equality. 
We allow all standard logical connectives and quantifiers in the language 
and, in addition, assume that it contains the logical constants $\top$, $\bot$ for true and false, respectively.
Without loss of generality, we restrict ourselves to closed formulas, i.e.\ we do not allow formulas to contain free variables.
The non-logical symbols of a formula $F$, denoted by $\lang{F}$, are all the predicate symbols and function symbols (including constants) occurring in $F$. 
Note that this excludes (quantified) variables and the equality symbol.

An \emph{axiomatisable theory}, or simply a \emph{theory} is any set of formulas.
For example, we can use the theory of linear integer arithmetic or the theory of lists.
We will from now on restrict ourself to a fixed theory $\mathcal{T}$ and give all definitions relative to $\mathcal{T}$.
This includes that we write $F_1, \dots, F_n \vDash F$ (instead of $F_1, \dots, F_n \vDash_\mathcal{T} F$) to denote 
that every model of $\mathcal{T}$ which satisfies each $F_1, \dots, F_n$ also satisfies $F$.

\begin{Definition}
Let $F_1, \dots, F_n, F$ be formulas, $n \geq 0$.
An 
\emph{inference rule} $R$ is a tuple $(F_1, \dots, F_n,F)$. 
An \emph{inference system} $\infsys$ is a set of inference rules. 

An inference rule $R = (F_1, \dots, F_n,F)$ is \emph{sound}, 
if $F_1, \dots, F_n \vDash F$.
An inference system $\infsys$ is called \emph{sound}, 
if it only consists of \emph{sound} inference rules.
\end{Definition}


From now on, we further restrict ourselves to a fixed 
inference system $\infsys$ which is sound (relative to $\mathcal{T}$)
and give all definitions relative to that system.

\paragraph{\bf Derivations and proofs.} 
We model logical proofs as directed hypergraphs in which 
vertices are associated with formulas and (hyper-)edges with inferences.
Because an inference always has exactly one conclusion, 
we only need hypergraphs where each edge has exactly one end vertex. 
%
%
Moreover, because the order of premises of an inference may be important, 
we use tuples to model the edges.
We will from now on refer to such (hyper-)edges simply as \emph{inferences}.


\begin{Definition}
Let $G$ be a formula and $\mathcal{F}$ a set of formulas.
A \emph{proof of $G$ from axioms $\mathcal{F}$} 
is a
finite 
acyclic 
labeled directed hypergraph $P = (V,E,L)$,
where $V$ is a set of vertices, $E$ a set of inferences,
and $L$ is a labelling function mapping each vertex $v \in V$ to a formula $L(v)$. 
For an inference $\infc{} \in E$ of the form $(v_1, \dots, v_n, v)$,
where $n \geq 0$, we call $v_1, \dots, v_n$ the \emph{premises of $\infc{}$}
and $v$ the \emph{conclusion of $\infc{}$}.

Additionally, we require the following:
\begin{enumerate}
\item \label{cond:exactly_one_inference}
	Each vertex $v \in V$ is a conclusion of exactly one inference $r \in E$.
\item
  There is exactly one vertex $v_0 \in V$ that is not a premise of any inference $\infc{} \in E$ and $L(v_0) = G$.
\item 
	each $\infc{} \in E$ is either
	\begin{enumerate*}
    \item an in\-fe\-rence of the form $(v)$ and $L(v) \in \mathcal{F}$, or
    \item an in\-fe\-rence of the form $(v_1, \dots, v_n, v)$ and $(L(v_1), \dots, L(v_n), L(v)) \in \infsys$.
  \end{enumerate*}
  In the first case, we call $\infc{}$ an \emph{axiom inference},
in the second case, $\infc{}$ is called a \emph{proper inference}.
\end{enumerate}
A \emph{refutation} from axioms $\mathcal{F}$ is a proof of the formula $\bot$ from $\mathcal{F}$.
\end{Definition}

Note that in order to support multiple occurrences of the same formula in a proof,
one needs to distinguish between vertices and the formulas assigned to them via the labelling function $L$.
However, because this generality is orthogonal to the ideas we want to present,
we will from now on identify each node $v \in V$ with its formula $L(v)$
and stop referring to the labelling function explicitly. 

In the above definition, condition \ref{cond:exactly_one_inference} ensures that any formula of the proof is justified by exactly one inference.
Later on we will look at subgraphs of a proof, which are not necessarily proofs themselves and in particular do not satisfy condition 1, since they contain formulas, which are not justified by any inference of the subgraph. We call such a subgraph a derivation 
and call the formulas which are not justified by any inference the premises of the derivation.
We can see a proof as a derivation having no premises.

\begin{Definition}
The definition of a \emph{derivation of $G$ from axioms $\mathcal{F}$} is the same 
as that of a proof $P = (V,E,L)$ of $G$ from $\mathcal{F}$,
except that condition \ref{cond:exactly_one_inference} is generalised to:
\begin{enumerate}
  \item  Each formula $F \in V$ is a conclusion of \emph{at most one} inference $\infc{} \in E$.
\end{enumerate}
The \emph{set of premises of a derivation $P$}, denoted by $\Prem(P)$, consists of all formulas $F \in V$, 
such that there exists no inference $r \in E$ with conclusion $F$.
\end{Definition}

The definition of a derivation is not natural as it distinguishes between axioms and premises. 
This distinction is, however, very important for us,
as it enables a succinct presentation of the results in Sect.~\ref{section:local_proof_interpolation}.

\begin{Lemma}[Soundness]
Let $P$ be a derivation of $G$ from axioms $\mathcal{F}$. Then we have
$$ \textstyle \mathcal{F} \vDash ( \bigwedge_{F_k \in \Prem(P)} F_k) \rightarrow G.$$
\end{Lemma}

To formalise the idea of a proof traversal in which the inferences are 
considered one by one from axioms to the final formula $G$, we make use of topological orderings.

\begin{Definition}
Let $P = (V,E,L)$ be a derivation. A topological ordering $<^T$ for $P$ is a linear ordering $<^T$ on $E$
such that for any two inferences $r_1,r_2 \in E$ if the conclusion of $r_1$ is a premise of $r_2$ then
$r_1 <^T r_2$.
\end{Definition}
A topological ordering exists for every derivation, because proofs, and thus also derivations, are required to be acyclic.

\paragraph{\bf Interpolation.} We now recall the notion of a logical
interpolant. 

\begin{Definition}
Let $A$ and $B$ be formulas.
\begin{enumerate}
  \item A non-logical symbol $s \in \lang{A \rightarrow B}$ is called \emph{$A$-local},
    if $s \in \lang{A} \setminus \lang{B}$,
    \emph{$B$-local}, 
    if $s \in \lang{B} \setminus \lang{A}$,
    and \emph{global} otherwise.
  \item   An \emph{interpolant} for $A,B$ is a formula $I$ such that
    $\vDash A \rightarrow I$, 
    $\vDash I \rightarrow B$ 
    and all non-logical symbols of $I$ are global.
\end{enumerate}
\end{Definition}
Craig's interpolation theorem~\cite{DBLP:journals/jsyml/Craig57} guarantees the existence 
of an interpolant for any pair of formulas $A, B$ for which $\vDash A \rightarrow B$.
In the sequel, we assume $A$ and $B$ to be
fixed and give all definitions relative to $A$ and $B$.

\paragraph{\bf Refutational theorem proving.} 
%
%
 To prove a first-order formula $F$ in practice, 
a refutational theorem prover proceeds by negating the input formula,
applying a normal form transformation, such as the Conjunctive Normal Form transformation, to the negation,
and deriving a contradiction $\bot$ from the obtained set of formulas $\mathcal{C}_{\neg F} = \mathit{CNF}(\neg F)$.
More specifically, in the case of proving the implication $A \rightarrow B$, the prover starts with axioms 
$\mathit{CNF}(A \land \neg B)$.

This is relevant for our work, because we rely on refutations as input for our method.
However, a complication arises, because the normal form transformations $\mathit{CNF}$ 
typically involves steps like sub-formula naming and Skolemisation \cite{DBLP:books/el/RV01/NonnengartW01,GCAI2016:New_Techniques_in_Clausal_Form_Generation},
which 1) introduce new non-logical symbols, 2) in general do not preserve logical equivalence.

To deal with 1) we impose a restriction on $\mathit{CNF}$ which dictates that the symbols
newly introduced on behalf of $A$ and $\neg B$ do not overlap. Formally, we require
\begin{equation} \label{reasonable_requirement}
\lang{A} \cap \lang{\neg B} = \lang{\mathit{CNF}(A)} \cap \lang{\mathit{CNF}(\neg B)},
\end{equation}
which is a very natural condition, because the newly introduced symbols are invariably required to be fresh.\footnote{
This could potentially be violated by an advanced transformation based on formula sharing. 
In particular, the case would need to involve a common sub-formula of $A$ and $\neg B$.}

To deal with 2), let us first recall that steps like sub-formula naming and Skolemisation,
although they do not preserve logical equivalence, do preserve satisfiability.
While this is sufficient to guarantee soundness of refutational theorem proving, 
it is not enough for the purposes of interpolation.
Fortunately, a stronger property, which is rarely stated explicitly, 
usually holds for the normal form transformation,
namely the
\emph{preservation of models over the common symbols}.
Formally, we require for every formula $F$ that 
\begin{itemize}
\item
every model $\mathcal{M}'$ of $\mathit{CNF}(F)$ is also a model of $F$, and
\item
every model $\mathcal{M}$ of $F$ can be extended to $\mathcal{M}'$ which is a model of $\mathit{CNF}(F)$,
\end{itemize}
where extended means that $\mathcal{M}'$ restricted to $\lang{F}$ equals $\mathcal{M}$.

Equipped with a transformation $\mathit{CNF}$ satisfying the above requirements, 
the general approach to interpolation from refutations consists of the following steps:
\begin{enumerate}
\item
	Given formulas $A$ and $B$, compute the respective normal forms 
	$\mathcal{C}_A = \mathit{CNF}(A)$ and $\mathcal{C}_{\neg B} = \mathit{CNF}(\neg B)$.
\item
	Find a refutation $P$ from axioms $\mathcal{C}_A\cup \mathcal{C}_{\neg B}$.
	
\item
	Extract from $P$ a formula $I$ such that 
	$\mathcal{C}_A \vDash I$, 
	$\mathcal{C}_{\neg B}, I \vDash \bot$, and all non-logical symbols of $I$ are global.\footnote{
Note that the symbols are global with respect to $A$ and $B$ if and only if they are global
with respect to $\mathcal{C}_A$ and $\mathcal{C}_{\neg B}$ thanks to the requirement \eqref{reasonable_requirement}.}
\end{enumerate}
\begin{Lemma}
The formula $I$ obtained in the last step is an interpolant for $A$ and $B$.
\end{Lemma}


%% file: local_proof_interpolation.tex
\section{Interpolants from refutations} 
\label{section:local_proof_interpolation}

We can separate the properties of an interpolant into two parts, the logical part and the restriction to the global symbols. Instead of considering only interpolants, we now want to look more generally at the formulas, which satisfy the logical part of the properties of interpolants, but not necessarily the restriction to the global symbols. We call such formulas intermediants.\footnote{
Bonacina and Johansson \cite{Bonacina15} introduce the notion of a \emph{provisional interpolant} with an analogous definition. However, the intended use of the notion is different. 
While provisional interpolants are meant to be modified to yield interpolants in a refinement stage, we give conditions under which intermediants are, in fact, interpolants.}

\begin{Definition}
Let $A, B$ be two formulas.
An \emph{intermediant} for $A,B$ is a formula $I$ such that we have both
    $\vDash A \rightarrow I$ and
    $\vDash I \rightarrow B$.
\end{Definition}

In the first part of this section, we want to investigate the space of intermediants,
which is induced by a given refutation. In the second part, we look at the subspace of those intermediants
which also respect the restriction on the global symbols, i.e.\ the formulas which are interpolants.

\subsection{Splitting refutations}

Let us now show how to use a refutation of $A \rightarrow B$ to construct intermediants. 
Intuitively, we want to split the refutation into two parts 
and construct a formula which describes the boundaries between the parts.

In the light of the discussion at the end of the previous section,
we assume the formulas $A$ and $\neg B$ have been transformed to sets of axioms
$\mathcal{C}_A$ and $\mathcal{C}_{\neg B}$.
It is also natural to extend the notion of an intermediant to axiom sets:

\begin{Definition}
Let $\mathcal{C}_A$ and $\mathcal{C}_{\neg B}$ be two sets of axioms.
An \emph{intermediant} for $\mathcal{C}_A,\mathcal{C}_{\neg B}$ is a formula $I$ such that we have both
    $\mathcal{C}_A \vDash I$ and
    $\mathcal{C}_{\neg B}, I  \vDash \bot$.
\end{Definition}

\begin{guide}
Splitting a proof into two parts for us means 
mapping each inference to one of the two parts.
Formally, we introduce a two element set $\{\partA, \partB\}$ to serve as a co-domain 
of such mapping, where $\partA$ denotes the $A$-part and $\partB$ the $B$-part. 
It is natural to map the axioms from $\mathcal{C}_A$ to $\partA$ and 
the axioms from $\mathcal{C}_{\neg B}$ to $\partB$, 
therefore we only consider mappings of this form. 
All other inferences can be mapped to any part.
\end{guide}

\begin{Definition}
Let $P$ be a refutation from axioms $\mathcal{C}_A \cup \mathcal{C}_{\neg B}$.
A \emph{splitting function} $\mathcal{S}$ is a function assigning each inference of $P$ to either $\partA$ or $\partB$, 
such that for each axiom inference $r = (F)$, if $\mathcal{S}(r) = \partA$ then $F \in \mathcal{C}_A$ and 
$r$ is called an \emph{$A$-axiom}, and 
if $\mathcal{S}(r) = \partB$ then $F \in \mathcal{C}_{\neg B}$ and $r$ is called a \emph{$B$-axiom}.
\end{Definition}

A given splitting function $\mathcal{S}$ splits a proof into several maximal subderivations. We now want to capture this intuitive notion formally.
We start with the concept of $\In$-formulas (resp. $\Out$-formulas) of $P$ and $\mathcal{S}$. Intuitively, these are the formulas which occur at the boundary between the subderivations.

\begin{Definition}
Let $P = (V,E,L)$ be a refutation from axioms $\mathcal{C}_A \cup \mathcal{C}_{\neg B}$ and let $\mathcal{S}$ be a splitting function on $P$.
The \emph{set of in-formulas}, 
which is denoted $\In(P,\mathcal{S})$, 
consists of those formulas $F \in V$, which has the following properties:
\begin{itemize}
	\item There exists an inference $r_1 \in E$ with conclusion $F$ and $\mathcal{S}(r_1) = \partB$.
 	\item There exists an inference $r_2 \in E$ with premise $F$ and $\mathcal{S}(r_2) = \partA$.
\end{itemize}
The \emph{set of out-formulas}, denoted $\Out(P,\mathcal{S})$, 
consists of formulas $F \in V$, such that
\begin{itemize}
	\item There exists an inference $r_1 \in E$ with conclusion $F$ and $\mathcal{S}(r_1) = \partA$.
 	\item Either there exists an inference $r_2 \in E$ with premise $F$ and $\mathcal{S}(r_2) = \partB$, or $F = \bot$.
\end{itemize}
\end{Definition}
Notice that the notions of in- and out-formulas are not entirely symmetrical. The reason for this will become clear later.

We are now able to formally introduce the maximal subderivations.

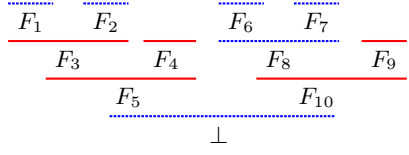
\begin{figure}[t]
\label{fig:ex_subderivation}
\begin{center}{}
    \input{pics/pic_ex_local_proof}
\end{center}
\caption{Consider the proof above along with the splitting function which is denoted by drawing the inferences assigned to $\partA$ using solid red lines and the inferences assigned to $\partB$ using dashed blue lines. The maximal $A$-subderivation of $F_5$ has premises $F_1$, $F_2$ (and conclusion $F_5$), the maximal $A$-subderivation of $F_{10}$ has premise $F_8$. The maximal $B$-subderivation of $F_1$ has no premises, the maximal $B$-subderivation of $\bot$ has premises $F_5$ and $F_{10}$. The $\In$-formulas are $F_1, F_2$ and $F_8$ and the $\Out$-formulas are $F_5$ and $F_{10}$. The induced simple splitting formula is $((F_1 \land F_2) \rightarrow F_5) \land (F_8 \rightarrow F_{10})$.}
\end{figure}

\begin{Definition}
Let $P = (V,E,L)$ be a refutation from axioms $\mathcal{C}_A \cup \mathcal{C}_{\neg B}$ and let $\mathcal{S}$ be a splitting function on $P$.
%
	Let $r \in E$ be an inference 
	and let $\{r_1, \dots, r_l\}$ be the set of those inferences
	which derive a premise of $r$ and are mapped by $\mathcal{S}$ to the same part as $r$,
	i.e. $\mathcal{S}(r)=\mathcal{S}(r_i)$ for $i=1,\ldots l$.
	Then we define $\Sub(r)$ recursively as
	$$\Sub(r) = \{ r \} \cup \Sub(r_1) \cup \dots \cup \Sub(r_l).$$
	Now let $F \in \Out(P,\mathcal{S})$ (resp.\ $F \in \In(P,\mathcal{S})$) be a formula and $r$ be the inference deriving $F$.
	We define the \emph{maximal $A$-subderivation (resp. $B$-subderivation) of $F$}, 
	denoted by $\Sub(F)$, as the induced derivation $(V', \Sub(r), L)$, 
	where $V'$ contains every vertex which is either a premise or a conclusion of an inference in $\Sub(F)$.
	We call $F$ the \emph{conclusion} of $\Sub(F)$.
	%

	The dependencies of $F$, written $\Dep(F)$, are defined as the premises of $\Sub(F)$.
\end{Definition}
We can observe that the $\In$-formulas (resp. $\Out$-formulas) are the premises (resp. conclusions) of all maximal $A$-subderivations. 
Dually, the $\In$-formulas (resp. $\Out$-formulas) are the conclusions (resp. premises) of all maximal $B$-subderivations.
The use of the introduced concepts is demonstrated in Fig.~\ref{fig:ex_subderivation}.

\setlength{\abovecaptionskip}{1pt}
\setlength{\belowcaptionskip}{1pt}

Note that the $A$-subderivations contain all $A$-axioms, but no $B$-axiom. 
Therefore the $A$-axioms's contribution to the derivation is captured by the $A$-subderivations. 
The key idea of this subsection is that encoding the contribution of the $A$-subderivations as a formula therefore yields the intermediant $I$ we are looking for.
The following lemma tells us how to describe the contribution of an $A$-subderivation.
\begin{Lemma}
\label{lemma:dependency_entailment}
Let $P$ be a refutation from axioms $\mathcal{C}_A \cup \mathcal{C}_{\neg B}$ and let $\mathcal{S}$ be a splitting function on $P$.
\begin{enumerate}
	\item 
	\label{lemma:dependency_entailment_1}
	Let $F \in \Out(P,\mathcal{S})$. Then we have $\mathcal{C}_A \vDash (\bigwedge_{F_k \in \Dep(F)} F_k) \rightarrow F$.
	\item 
	\label{lemma:dependency_entailment_2}
	Let $F \in \In(P,\mathcal{S})$. Then we have $\mathcal{C}_{\neg B} \vDash (\bigwedge_{F_k \in \Dep(F)} F_k) \rightarrow F$.
\end{enumerate}
\end{Lemma}

We therefore arrive at the following definition.
\begin{Definition}
\label{def:simple_interpolant}
Let $P$ be a refutation from axioms $\mathcal{C}_A \cup \mathcal{C}_{\neg B}$ and let $\mathcal{S}$ be a splitting function on $P$.
The formula $$ I := \bigwedge_{F \in \Out(P,\mathcal{S})} ( (\bigwedge_{F_k \in \Dep(F)} F_k) \rightarrow F  )$$
is called the \emph{simple splitting formula of $P$ induced by $\mathcal{S}$}.
\end{Definition}

\begin{Theorem}
\label{theorem:splitting_formula_entailment}
Let $P$ be a refutation from axioms $\mathcal{C}_A \cup \mathcal{C}_{\neg B}$ and let $\mathcal{S}$ be a splitting function on $P$.
Then the simple splitting formula $I$ induced by $\mathcal{S}$ is an intermediant. 
\end{Theorem}

\begin{Proof}
~
\begin{enumerate}
	\item 
    For each $F \in \Out(P,\mathcal{S})$, we can use Lemma \ref{lemma:dependency_entailment}.\ref{lemma:dependency_entailment_1}
    to get $\mathcal{C}_A \vDash (\bigwedge_{F_k \in \Dep(F)} F_k) \rightarrow F$.
    Therefore we have $\mathcal{C}_A \vDash \bigwedge_{F \in \Out(P,\mathcal{S})} ((\bigwedge_{F_k \in \Dep(F)} F_k) \rightarrow F)$.
	\item
    Let $<^T$ be a topological ordering for $P$
	and let $F_1, \dots, F_n$ denote the formulas of $\In(P) \cup \Out(P)$ in the order induced by $<^T$.
	We visit the formulas from $F_1$ to $F_n$
	and prove by complete induction that 
	$I, \mathcal{C}_{\neg B} \vDash F_1, \dots, F_i$.
	Since $F_n = \bot$, we afterwards are able to conclude $I, \mathcal{C}_{\neg B} \vDash \bot$.

    Inductive step:
    Let us assume, by the induction hypothesis, that $I, \mathcal{C}_{\neg B} \vDash F_1, \dots, F_{i-1}$. 
    We make a case distinction on $\mathcal{S}(r)$, where $r$ is the inference which derived $F_i$:
    \begin{itemize}
        \item Case $\mathcal{S}(r) = \partA$: 		
		By the definition of $I$, we know that 
        $I \vDash (\bigwedge_{F_k \in Dep(F_i)} F_k) \rightarrow F_i$.
		Using both the definition of topological orderings and the definition of $\Dep$ we know that 
		$\Dep(F_i) \subseteq \{ F_1, \dots F_{i-1}\}$,
		so we can combine the previous facts 
		to obtain $I, \mathcal{C}_{\neg B} \vDash F_1, \dots, F_i$. 
        \item Case $\mathcal{S}(r) = \partB$:       
        We use Lemma \ref{lemma:dependency_entailment}.\ref{lemma:dependency_entailment_2}
        to conclude $\mathcal{C}_{\neg B} \vDash (\bigwedge_{F_k \in \Dep(F_i)} F_k) \rightarrow F_i$. 
        As in the previous case, we can use $\Dep(F_i) \subseteq \{ F_1, \dots F_{i-1}\}$ 
        to conclude $I, \mathcal{C}_{\neg B} \vDash F_1, \dots, F_i$.
    \end{itemize}
\end{enumerate} 
\end{Proof}
We summarise the ideas of this subsection in \simpleAlg{} (Algorithm~\ref{main_alg}).

\begin{algorithm}
	\caption{\simpleAlg{}}
 	\label{main_alg}
	\begin{algorithmic}
		\State choose a splitting function on $P$.
		\State compute $\Out(P)$ and $\Dep(F)$ for all $F$ using depth first search
		\State \Return $I$ as defined in Definition \ref{def:simple_interpolant}
	\end{algorithmic}
\end{algorithm}

\subsection{Intermediants of linear size}
\simpleAlg{} yields an intermediant of size which is in the worst case quad\-ra\-tic in the size of the proof. 
This may be prohibitively large for large proofs.
In this subsection, we describe
an algorithm which yields intermediants of size which is linear in the size of the proof. 
Modifying Algorithm \ref{main_alg} to generate such an intermediant is nontrivial: 
there are examples, where the simple splitting formula
is provably logically stronger than any intermediant which uses every formula of the refutation only once, 
cf. Fig.~\ref{fig:ex_weakening_linear}.
We therefore need to modify the algorithm 
such that it produces an intermediant
which is logically weaker but still sufficiently strong to be inconsistent with $C_B$. 

The key idea for the new algorithm is contained in the following definition.

\begin{figure}[t]
\begin{center}
    \input{pics/ex_weakening_linear}
\end{center}
\caption{
	Let $r_1 = (F_1,F_3)$, $r_2 = (F_2,F_4)$, 
	$r_3 = (F_1, F_2, F_5)$, 
	and $r_4 = (F_3, F_4, F_5, \bot)$. 
	Let further $\mathcal{S}(r_1) = \mathcal{S}(r_2) = \mathcal{S}(r_3) = \partA$ 
	and $\mathcal{S}(r_4) = \partB$. 
	Then Algorithm \ref{main_alg} generates the simple splitting formula 
	$I = (F_1 \rightarrow F_3) \land (F_2 \rightarrow F_4) \land ((F_1 \land F_2) \rightarrow F_5)$. 
	There is no intermediant
	which is both logically equivalent to $I$ and contains each formula of the given proof at most once.
}
\label{fig:ex_weakening_linear}
\end{figure}

\begin{Definition}
\label{def:linear_interpolant}
Let $P$ be a refutation from axioms $\mathcal{C}_A \cup \mathcal{C}_{\neg B}$, 
$\mathcal{S}$ a splitting function on $P$,
and let $<^T$ be a topological ordering for $P$. 
Furthermore let $F_1, \dots, F_n$ denote the formulas of $\In(P) \cup \Out(P)$ ordered by $<^T$.
Now let 
$$I_i = 
\begin{cases} 
\top 						& \mbox{if } i = n+1 \\ 
F_i \rightarrow I_{i+1}		& \mbox{if } F_i \in \In(P) \\ 
F_i \land I_{i+1}			& \mbox{if } F_i \in \Out(P) \\ 
\end{cases}$$
Then $I_1$ is called \emph{linear splitting formula of $P$ induced by $\mathcal{S}$ and $<^T$}.
\end{Definition}

\begin{guide}
Note that the size of $I_1$ is linear in the size of $P$ in Definition \ref{def:linear_interpolant}.  
\end{guide}

\begin{Theorem}
Let $P$ be a refutation from axioms $\mathcal{C}_A \cup \mathcal{C}_{\neg B}$, 
let $\mathcal{S}$ be a splitting function on $P$ 
and let $<^T$ be a topological ordering for $P$. 
Then the linear splitting formula $I$ induced by $\mathcal{S}$ and $<^T$ is an intermediant.
\end{Theorem}

\begin{Proof}
Let
$$I' = \bigwedge_{F_i \in \Out(P)} ((\bigwedge_{F_k \in \In(P), F_k <^T F_i} F_k) \rightarrow F_i).$$
First note that $I'$ is logically equivalent to $I$: This can be proved by a simple induction using the two facts that conjunction on the right distributes over implication and that $A \rightarrow (B \rightarrow C)$ is equivalent to $(A \land B) \rightarrow C$.

Now we complete the proof by showing that $I'$ is an intermediant:

\begin{enumerate}
    \item 
    Using both the definition of topological orderings and the definition of $\Dep$ we know that 
	$\Dep(F_i) \subseteq \{F_k\in \In(P) \mid  F_k <^T F_i\}$,
	so $I'$
	is logically weaker than the simple splitting formula. 
	Therefore $\mathcal{C}_A \vDash I'$ follows from Theorem~\ref{theorem:splitting_formula_entailment}.1.
    \item
    We can show $\mathcal{C}_{\neg B}, I' \vDash \bot$ by re-using the proof of Theorem \ref{theorem:splitting_formula_entailment}.2 
    with $<^T$ as the topological ordering and by replacing $\Dep(P)$ with $\{F_k\in \In(P) \mid  F_k <^T F_i\}$.
\end{enumerate} 
\end{Proof}
We summarise the presented ideas in \linearAlg{} (Algorithm \ref{linear_alg})
and conclude this subsection by pointing out the following basic lemma, which will become useful later in the paper.

\begin{algorithm}[t]
	\caption{\linearAlg{}}
	\label{linear_alg}
	\begin{algorithmic}
		\State choose a splitting function and a topological ordering on $P$.
		\State compute $\In(P)$ and $\Out(P)$
		\State \Return $I_1$ as defined in Definition \ref{def:linear_interpolant}
	\end{algorithmic}
\end{algorithm}

\begin{Lemma}
\label{lemma:different-parts}
Let $P = (V,E,L)$ be a refutation from axioms $\mathcal{C}_A \cup \mathcal{C}_{\neg B}$ and let $\mathcal{S}$ be a splitting function on $P$. 
Let further $I$ be the linear splitting formula induced by $\mathcal{S}$ and 
let $F \in V$ be an arbitrary formula different from $\bot$.
%
Then $F$ occurs in $I$ if and only if there are two inferences $r_1,r_2$, where $r_1$ derives $F$, $F$ is a premise of $r_2$ and $\mathcal{S}(r_1) \neq \mathcal{S}(r_2)$.
\end{Lemma}

\subsection{Interpolants as special intermediants}

In the previous subsections, we discussed how to construct intermediants given a splitting function.
We now look closer at the question which splitting function to choose. 
While studying the intermediants induced by different choices of a splitting function is an interesting topic in general,
we turn our attention to the problem of choosing a splitting function
such that the induced intermediant is an interpolant,
i.e.\ we have the additional requirement that the intermediant contains no local symbols.

Let us recall the notion of local proofs---also called split proofs---introduced by Jhala and McMillan~\cite{jm06}:

\begin{Definition}[Local Proof]
\label{def:localproof}
A proof $P=(V,E,L)$ from axioms $\mathcal{C}_A \cup \mathcal{C}_{\neg B}$
is \emph{local} if for every inference $(F_1, \ldots, F_n,F) \in E$
we have either: 
\begin{itemize}
\item $\lang{F_1}\cup \ldots \cup \lang{F_k} \cup \lang{F} \subseteq \lang{\mathcal{C}_A}$ or
\item $\lang{F_1}\cup \ldots \cup \lang{F_k} \cup \lang{F} \subseteq \lang{\mathcal{C}_{\neg B}}$.
\end{itemize}
\end{Definition}
The definition of local proofs ensures that we can define a splitting function $\mathcal{S}$
which maps all inferences with $A$-local symbols to $\partA$ and those with $B$-local symbols to $\partB$.

\begin{Definition}
Let $P$ be a local proof. 
A \emph{local splitting function on $P$} is a splitting function $\mathcal{S}$ on $P$ 
such that $\mathcal{S}(r) = \partA$ (resp. $\mathcal{S}(r) = \partB$) for all inferences $r$ 
having as premise or conclusion a formula containing an $A$-local (resp. a $B$-local) symbol.
\end{Definition}

\begin{guide}
The corollary of the following lemma represents the central observation of this subsection: 
local proofs are exactly the proofs on which we can define a splitting function
that induces an intermediant which is an interpolant.
\end{guide}

\begin{Lemma}
\label{lemma:splitting_formula_global}
Let $P = (V,E,L)$ be a refutation from axioms $C_A \cup C_B$, 
$\mathcal{S}$ be a local splitting function on $P$,
and $I$ the corresponding simple (resp. linear) splitting formula.
\begin{enumerate}[label=\roman*)]
    \item \label{lemma:splitting_formula_global_1} 
    Then any formula $F \in \In(P, \mathcal{S}) \cup \Out(P, \mathcal{S})$ contains neither an $A$-local nor a $B$-local symbol.
    \item \label{lemma:splitting_formula_global_2} $I$ contains neither $A$-local nor $B$-local symbols.
\end{enumerate}
\end{Lemma}

\begin{Proof}
\begin{enumerate}[label=\roman*)]
    \item 
    Consider any formula $F \in \Out(P, \mathcal{S})$.
    If $F = \bot$ then $F$ trivially contains neither an $A$-local nor a $B$-local symbol.
    Otherwise, we know that
    there exists an inference $r_1 \in E$ with premise $F$ and $\mathcal{S}(r_1) = \partB$.
    By the locality of $\mathcal{S}$ we get that $F$ contains no $A$-local symbol.
    Furthermore, we know that there exists an inference $r_2 \in E$ with conclusion $F$ and $\mathcal{S}(r_2) = \partA$.
    By the locality of $\mathcal{S}$ we get that $F$ contains no $B$-local symbol.
    
    Now consider any formula $F \in \In(P, \mathcal{S})$. 
    We can use a similar argument to show that $F$ contains neither an $A$-local nor a $B$-local symbol.

    \item Follows immediately from \ref{lemma:splitting_formula_global_1} and the definition of the simple (resp. linear) splitting formula.
\end{enumerate}
\end{Proof}

\begin{Corollary}
\label{corollary:main}
Let $P$ be a local refutation, let $\mathcal{S}$ be a local splitting function on $P$ and let $I$ 
be either the simple splitting formula or the linear splitting formula.
Then $I$ is an interpolant for $A,B$.
\end{Corollary}

%% file: pics/pic_ex_local_proof.tex
	\begin{tikzpicture}[yscale=0.5]
        \tikzset{partB/.style={thick, blue, dash pattern= on 1pt off 0.5pt,dash phase= 0pt}}

        \draw [white] (-3,-0.7) rectangle (2.7,4);

        \node (bot)     at (0,0) {$\bot$};
        
        \node (f5)      at ($(bot)+(-1.2,1)$) {$F_5$};
        \node (f10)      at ($(bot)+(1.3,1)$) {$F_{10}$};
        \draw [partB] ($(f5)+(-0.25,-0.5)$) -- ($(f10)+(0.25,-0.5)$);

        \node (f3)      at ($(f5)+(-0.8,1)$) {$F_3$};
        \node (f4)      at ($(f5)+(0.5,1)$) {$F_4$};
        \draw [red, thick] ($(f3)+(-0.3,-0.5)$) -- ($(f4)+(0.4,-0.5)$);    
        \draw [red, thick] ($(f4)+(-0.3,+0.5)$) -- ($(f4)+(0.4,+0.5)$);    

        \node (f1)      at ($(f3)+(-0.5,1)$) {$F_1$};
        \node (f2)      at ($(f3)+(0.5,1)$) {$F_2$};
        \draw [red, thick] ($(f1)+(-0.3,-0.5)$) -- ($(f2)+(0.3,-0.5)$);
        \draw [partB] ($(f1)+(-0.3,+0.5)$) -- ($(f1)+(0.3,+0.5)$);
        \draw [partB] ($(f2)+(-0.3,+0.5)$) -- ($(f2)+(0.3,+0.5)$);

        \node (f8)      at ($(f10)+(-0.5,1)$) {$F_8$};
        \node (f9)      at ($(f10)+(0.9,1)$) {$F_{9}$};
        \draw [red, thick] ($(f8)+(-0.3,-0.5)$) -- ($(f9)+(0.3,-0.5)$);           
        \draw [red, thick] ($(f9)+(-0.3,0.5)$) -- ($(f9)+(0.3,0.5)$);   

        \node (f6)      at ($(f8)+(-0.5,1)$) {$F_6$};
        \node (f7)      at ($(f8)+(0.5,1)$) {$F_7$};
        \draw [partB] ($(f6)+(-0.3,-0.5)$) -- ($(f7)+(0.3,-0.5)$);
        \draw [partB] ($(f6)+(-0.3,+0.5)$) -- ($(f6)+(0.3,+0.5)$);
        \draw [partB] ($(f7)+(-0.3,+0.5)$) -- ($(f7)+(0.3,+0.5)$);                
	\end{tikzpicture}

%% file: pics/ex_weakening_linear.tex
\begin{tikzpicture}[yscale=0.5]


        \node (bot)     at (0,0) {$\bot$};

        \node (f3)      at ($(bot)+(-1,1)$) {$F_3$};
        \node (f5)      at ($(bot)+(0,1)$) {$F_5$};
        \node (f4)      at ($(bot)+(1,1)$) {$F_4$};
        \draw [blue, thick] ($(f3)+(-0.25,-0.5)$) -- ($(f4)+(0.25,-0.5)$);          
        \node           at ($(f3)+(-0.4,-0.5)$) {$r_4$};

        \node (f1)      at ($(f3)+(0.3,1)$) {$F_1$};
        \node (f2)      at ($(f4)+(-0.3,1)$) {$F_2$};
        \draw [red, thick] ($(f3)+(-0.25,+0.4)$) -- ($(f3)+(0.35,+0.4)$);    
        \node           at ($(f3)+(-0.4,0.5)$) {$r_1$};
        \draw [red, thick] ($(f4)+(-0.35,+0.4)$) -- ($(f4)+(0.25,+0.4)$);    
        \node           at ($(f4)+(0.45,0.5)$) {$r_2$};
        \draw [red, thick] ($(f1)+(-0.1,-0.5)$) -- ($(f2)+(0.1,-0.5)$);          
        \draw [red, thick] ($(f1)+(-0.25,0.5)$) -- ($(f1)+(0.25,0.5)$);          
        \draw [red, thick] ($(f2)+(-0.25,0.5)$) -- ($(f2)+(0.25,0.5)$);          
\end{tikzpicture}

%% file: splitting_functions.tex
\section{Implementing Local Splitting Functions}
\label{section:splitting_functions}

\begin{guide}
By the definition of a local splitting function we know that 
we need to assign axioms and inferences with local symbols to the corresponding part. 
All the other inferences---%
the inferences forming the so called \emph{grey area} 
\cite{HoderKV12}---%
can be assigned freely to either part.
Different choices on how to split the grey area result in different $A$-subproofs
and therefore in different interpolants, which vary,
e.g., in size, the number of contained quantifiers and in logical strength.

We want to minimize the interpolant with respect to a given weight function $w$, 
which maps each formula $F$ to its weight $w(F)$. 
%
The task we want to solve in this section is, therefore, 
to be able to come up with a local splitting function
which minimises the weight of the resulting interpolant.

We present two different solutions, 
a heuristical greedy approach and one of expressing the optimal splitting as a minimisation problem.
Both solutions are based on the insight from Lemma~\ref{lemma:different-parts} of Sect.~\ref{section:local_proof_interpolation}: 
A conclusion $F$ of an inference $r_1$ occurs in the linear splitting formula 
if and only if there is an inference $r_2$ with $F$ as a premise 
such that the splitting function maps $r_1$ and $r_2$ to different parts.
\end{guide}

\subsection{Greedy weighted sum heuristic}

Consider an inference $r$ of the grey area with premises $C_1, \dots C_n, D_1, \dots, D_m$ 
and assume that the inferences deriving $C_1, \dots, C_n$ are already assigned to $\partA$ 
and that the inferences deriving $D_1, \dots, D_m$ are already assigned to $\partB$.
Using Lemma \ref{lemma:different-parts}, we know that if we assign $r$ to $\partA$, 
then $D_1, \dots, D_m$ will be added to the interpolant 
and if we assign $r$ to $\partB$, 
then $C_1, \dots, C_n$ will be added to the interpolant. 

We can therefore use the following greedy strategy to locally minimize the weight of the interpolant: 
for any inference $r$ of the grey area, if $\sum_{k=1}^n w(C_k) > \sum_{k=1}^m w(D_k)$, 
map $r$ to $\partA$, otherwise to $\partB$.
 
This results in \splitByHeur{} (Algorithm \ref{alg:top-down-weighted-sum-heuristic}):

\begin{algorithm}
	\caption{\splitByHeur{}}
	\label{alg:top-down-weighted-sum-heuristic}
	\begin{algorithmic}
		\For{each inference $r$ of $P$ (top-down)}
			\If{$r$ is an $A$-axiom or $r$ contains an $A$-local symbol}
				\State set $\mathcal{S}(r)$ to $\partA$
			\ElsIf{$i$ is a $B$-axiom or $r$ contains a $B$-local symbol}
				\State set $\mathcal{S}(r)$ to $\partB$
			\Else
				\If{$\sum_{k=1}^n w(C_k) > \sum_{k=1}^m w(D_k)$}
					\State set $\mathcal{S}(r)$ to $\partA$
				\Else
					\State set $\mathcal{S}(r)$ to $\partB$
				\EndIf
			\EndIf
		\EndFor
		\State\Return $\mathcal{S}$
	\end{algorithmic}
\end{algorithm}

\begin{figure}[t]
\begin{center}
    \input{pics/ex_heuristic_increasing}
\end{center}
\caption{
		Let $\mathcal{S}(i_1) = \partA$ and $\mathcal{S}(i_3) = \partB$. Let further $w(F_1) < w(F_2)$. Algorithm \ref{alg:top-down-weighted-sum-heuristic} would now assign $i_2$ to  $\partA$ and therefore include $F_2$ in the interpolant. It would be better to assign $i_2$ to $\partB$ in order to include $F_1$ in the interpolant instead of $F_2$.
	}
\label{fig:ex_heuristic_increasing}
\end{figure}
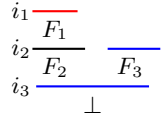

The two reasons why a locally optimal choice is not a globally optimal choice can be seen in Figures \ref{fig:ex_heuristic_dag} and \ref{fig:ex_heuristic_increasing}.



\begin{figure}[t]
\begin{center}
    \input{pics/ex_heuristic_dag}
\end{center}
\caption{
	Let $\mathcal{S}(i_1) = \mathcal{S}(i_3) = \partA$ and $\mathcal{S}(i_2)= \partB$. Let further $w(F_1) = w(F_3) = 2$ and $w(F_2) = 3$. For both inferences $i_4$ and $i_5$, 
	the assignment of the inference to $\partB$ is locally optimal, 
	then causes the assignment of $i_6$ to $\partB$ and finally yields an interpolant of weight $4$. 
	Note that $F_2$ is used as a premise of both $i_4$ and $i_5$,  
	so due to the DAG-structure we would only include it once
	if we assigned both $i_4$ and $i_5$ to $\partA$. 
	This would then cause the assignment of $i_6$ to $\partA$ and finally yield a smaller interpolant of weight 3.
	}
\label{fig:ex_heuristic_dag}
\end{figure}
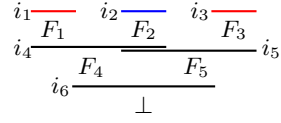

\subsection{Encoding optimal splitting as a minimisation problem}
Similar to the idea presented in \cite{HoderKV12}, we can alternatively 
encode the problem of finding an optimal local splitting function 
as a minimisation problem and pass it to a pseudo-boolean constraint solver.
This yields an optimal assignment, 
but is computationally more expensive.

The encoding works as follows.
We use propositional variables $x_i$ to denote that 
inference $i$ is assigned to $\partA$ 
and use propositional variables $L_i$ to denote that 
the conclusion of $i$ occurs in the interpolant. 
We again predict the size of the resulting interpolant using Lemma \ref{lemma:different-parts}, but this time use the optimisation procedure to make globally optimal choices instead of greedily making locally optimal ones. This leads to algorithm \splitOptimally{} (Algorithm~\ref{alg:weighted-sum-optimal}).

\begin{algorithm}
	\caption{\splitOptimally{}}
	\label{alg:weighted-sum-optimal}
	\begin{algorithmic}
		\For{each inference $r$ of $P$}
			\If{$r$ is an $A$-axiom or $r$ contains an $A$-local symbol}
				\State assert $x_r$
			\ElsIf{$r$ is a $B$-axiom or $r$ contains a $B$-local symbol}
				\State assert $\neg x_r$
			\EndIf
			\For{each parent inference $r'$ of $r$}
				\State assert $(\neg(x_r \leftrightarrow x_{r'})) \rightarrow L_{r'}$
			\EndFor
		\EndFor

		\State compute model $M$ which minimises $\sum_{r \in P} w(concl(r)) \cdot L_r$
		
		\For{each inference $r$ of $P$}
			\If{$x_r$ evaluates to true in $M$}
				\State set $\mathcal{S}(r)$ to $\partA$
			\Else
				\State set $\mathcal{S}(r)$ to $\partB$
			\EndIf
		\EndFor
		
		\State\Return $\mathcal{S}$
	\end{algorithmic}
\end{algorithm}

%% file: pics/ex_heuristic_dag.tex
\begin{tikzpicture}[yscale=0.5]


        \node (bot)     at (0,0) {$\bot$};

        \node (f3)      at ($(bot)+(-0.7,1)$) {$F_4$};
        \node (f5)      at ($(bot)+(0.7,1)$) {$F_5$};
        \draw [thick] ($(f3)+(-0.25,-0.5)$) -- ($(f5)+(0.25,-0.5)$);
        \node           at ($(bot)+(-1.1,0.5)$) {$i_6$};

        \node (f1)      at ($(f3)+(-0.5,1)$) {$F_1$};
        \node (f2)      at ($(f3)+(0.7,1)$) {$F_2$};
        \draw [thick] ($(f1)+(-0.3,-0.45)$) -- ($(f2)+(0.3,-0.45)$);    
        \node           at ($(f3)+(-0.9,0.5)$) {$i_4$};
        \draw [red, thick] ($(f1)+(-0.3,0.5)$) -- ($(f1)+(0.3,+0.5)$);
        \node           at ($(f1)+(-0.4,0.5)$) {$i_1$};
        \draw [blue, thick] ($(f2)+(-0.3,+0.5)$) -- ($(f2)+(0.3,+0.5)$);
        \node           at ($(f2)+(-0.45,0.5)$) {$i_2$};

        \node (f4)      at ($(f5)+(0.5,1)$) {$F_3$};
        \draw [thick] ($(f2)+(-0.3,-0.55)$) -- ($(f4)+(0.3,-0.55)$);    
        \node           at ($(f5)+(1,0.5)$) {$i_5$};
        \draw [red, thick] ($(f4)+(-0.3,0.5)$) -- ($(f4)+(0.3,+0.5)$);
        \node           at ($(f4)+(-0.45,0.5)$) {$i_3$};
\end{tikzpicture}

%% file: pics/ex_heuristic_increasing.tex
\begin{tikzpicture}[yscale=0.5]


        \node (bot)     at (0,0) {$\bot$};

        \node (f2)      at ($(bot)+(-0.5,1)$) {$F_2$};
        \node (f3)      at ($(bot)+(0.5,1)$) {$F_3$};
        \draw [blue, thick] ($(f2)+(-0.25,-0.5)$) -- ($(f3)+(0.25,-0.5)$);
        \node           at ($(bot)+(-0.95,0.5)$) {$i_3$};
        \draw [blue, thick] ($(f3)+(-0.3,+0.5)$) -- ($(f3)+(0.4,+0.5)$);    
        \node           at ($(f2)+(-0.45,0.5)$) {$i_2$};

        \node (f1)      at ($(f2)+(0,1)$) {$F_1$};
        \draw [thick] ($(f1)+(-0.3,-0.5)$) -- ($(f1)+(0.4,-0.5)$);    
        \node           at ($(f1)+(-0.45,0.5)$) {$i_1$};
        \draw [red, thick] ($(f1)+(-0.3,+0.5)$) -- ($(f1)+(0.3,+0.5)$);

\end{tikzpicture}

%% file: related.tex
\section{Discussion and Related Work}
\label{sec:related}

There are two main existing approaches to constructing 
interpolants from arbitrary local proofs 
in arbitrary sound first-order proof systems with equality.

First, there is the work from Jhala and McMillan (Theorem 3 of ~\cite{jm06}).
They present an algorithm which consists of two main phases: 
A) Extract a propositionally unsatisfiable set of formulas $F$, 
B) obtain a propositional refutation of $F$ using boolean constraint propagation 
   and apply McMillan's interpolation algorithm for propositional logic \cite{McMillan03} to the result 
   in order to obtain an interpolant for the original local refutation. 

One can easily see that the set $F$ constructed in phase A consists of both the conjuncts of the splitting formula from Definition~\ref{def:simple_interpolant} and the conjuncts of the simple splitting formula obtained by swapping $A$ and $B$ in the proof. In contrast, Algorithm~\ref{main_alg} only needs the former conjuncts. Furthermore we know from Corollary~\ref{corollary:main} that it is sufficient to conjoin all these conjuncts instead of unnecessarily constructing and interpolating from a propositional refutation. Besides conceptually simplifying the algorithm, this also enables the optimisations 
presented in Sect.~\ref{section:splitting_functions}.

More importantly, in \cite{McMillan08}, it is claimed that the complexity of the algorithm behind Theorem 3 of \cite{jm06} is linear in the size of the proof. 
While phase B of the algorithm is clearly linear, 
we can see easily from Example \ref{ex:mcmillan_quadratic} below that phase A is worst case quadratic in the size of the proof,
making the whole algorithm quadratic,
which is contrasts to out Algorithm~\ref{linear_alg}, that is linear.
\begin{Example}
\label{ex:mcmillan_quadratic}
Consider a split refutation with nodes $A_1, \dots ,A_n, B_1,\dots ,B_n, \bot$; 
edges $(A_i, A_{i+1}), (A_i, B_{i+1}), (B_i, A_{i+1}), (B_i, B_{i+1})$, for $1 \leq i < n$ and $(A_n, \bot)$, $(B_n, \bot)$;
and labeling $P(A_i) = A$, $P(B_i) = B$, $P(\bot) =$ arbitrary.
Phase A) would construct a graph with edges
$(A_i, B_j), (B_i, A_j)$ forall $0 < i < j \leq n$,
which is quadratic in $n$.
\end{Example}
The second main approach to constructing interpolants in first-order logic
with equality using an arbitrary sound inference system
was introduced in~\cite{DBLP:conf/cade/KovacsV09}
and later improved by an optimisation technique in~\cite{HoderKV12}.
Let us refer to the interpolation algorithm from \cite{DBLP:conf/cade/KovacsV09} as $\mathcal{SE}$.
In a nutshell, $\mathcal{SE}$ uses two main concepts:

As a first concept, it constructs the largest subderivations containing only symbols from one of the two partitions (cf. Lemma 8 of \cite{DBLP:conf/cade/KovacsV09}).
This construction corresponds to a commitment to a specific choice of local splitting function in our framework. 
In contrast, both Algorithm \ref{main_alg} and Algorithm \ref{linear_alg} are parametrized by an arbitrary local splitting function
and different choices yield different interpolants.

As the main contribution of \cite{HoderKV12}, the authors extend algorithm $\mathcal{SE}$
such that it also considers a space of different interpolants and optimise over this space. 
We can see that the extension simulates different choices of splitting function by merging proof steps. 
Both the algorithm from \cite{HoderKV12} and our Algorithm~\ref{alg:weighted-sum-optimal} encode the space of candidates and the minimisation objective as a pseudo-boolean constraint problem 
and then ask an optimising SMT-solver for an optimal solution.
While encoding the space of splitting functions is trivial using Algorithm \ref{alg:weighted-sum-optimal}, 
encoding the space of local proofs, which are results from repeated pairwise merging of inferences, 
is much more involved. 
More critically, 
while we can make use of Lemma~\ref{lemma:different-parts} to predict the size of the resulting interpolant, 
the approach from \cite{HoderKV12} uses a notion of so called digest 
to predict the size of the interpolant computed from the transformed proof. 
The authors claim that the interpolant is a boolean combination of formulas in the digest (Theorem 3.6, \cite{HoderKV12}). 
Unfortunately, this claim is wrong, which can be concluded from the counterexample presented in Fig.~\ref{fig:digest_counterexample}.
Therefore the technique presented in \cite{HoderKV12} can potentially yield sub-optimal interpolants.

\begin{figure}[t]
\begin{center}
    \input{pics/digest_counterexample}
\end{center}
\caption{Consider the proof above, taken from Example 5.2 in \cite{HoderKV12}. Let $R_1, R_3$ and $R_4$ be formulas containing $A$-local symbols, $B_1$ a formula containing $B$-local symbols and let $G_1, G_2, G_3$ and $G_6$ be formulas containing no local symbols. Then the digest contains only $G_6$, but the 
algorithm from \cite{DBLP:conf/cade/KovacsV09} would construct the interpolant $G_3 \land \neg G_6$, which also contains $G_3$.}
\label{fig:digest_counterexample}
\end{figure}

As the second concept, the algorithm $\mathcal{SE}$ from \cite{DBLP:conf/cade/KovacsV09} 
relies on a recursive construction to compute the interpolant: it computes for each largest subderivation a formula
such that the formula of the outermost call yields an interpolant (cf. Lemma 10 of \cite{DBLP:conf/cade/KovacsV09}). 
We now want to hint at the relation of algorithm $\mathcal{SE}$ and Algorithm \ref{main_alg}.
Consider a subderivation with premises $F_1, \dots, F_k$ and conclusion $F$. Let further $I_1, \dots, I_k$ denote the recursively computed formulas. Algorithm $\mathcal{SE}$ now constructs the following formulas:
\begin{itemize}
	\item Case $A$: $I = ((I_1 \lor F_1) \land \dots \land (I_k \lor F_k)) \land \neg(F_1 \land \dots \land F_k)$.
	\item Case $B$: $I = ((I_1 \lor F_1) \land \dots \land (I_k \lor F_k))$.
\end{itemize}
It is not difficult to see that one can reformulate the construction of $\mathcal{SE}$ as the following one, which we will refer to as $\mathcal{SE}'$:
\begin{itemize}
	\item Case $A$: $I = ((I_1 \lor F_1) \land \dots \land (I_k \lor F_k)) \lor F \land ((F_1 \land \dots \land F_k) \rightarrow F)$.
	\item Case $B$: $I = (I_1 \land \dots \land I_k)$.
\end{itemize}
Note that although the intermediate formulas of algorithm $\mathcal{SE}$ and $\mathcal{SE}'$ are potentially different, the result of the outermost call is the same for $\mathcal{SE}$ and $\mathcal{SE}'$.

We now state a recursive presentation of our Algorithm \ref{main_alg} in order to compare it to $\mathcal{SE}'$. 
The idea is to replace the global view on the refutation, i.e. the iteration over all elements of $\Out(P, \mathcal{S})$, by a recursive construction which collects all the formulas describing the boundaries of maximal $A$-subderivations.

Let $P$ be a local proof of a formula $F$ and let $r$ be the inference which derives $F$.
Let further $\mathcal{S}$ be a local splitting function on $P$.
We compute a formula using the following recursive construction:
Let $F_1, \dots, F_k$ denote the elements of $\Dep(F)$ and let $I_i$ denote the formula computed recursively from $F_i$.

\begin{itemize}
	\item Case $\mathcal{S}(r) = \partA$:
	$I = (I_1 \land \dots \land I_n) \land ((F_1 \land \dots \land F_n) \rightarrow F)$.
	\item Case $\mathcal{S}(e) = \partB$:
	$I = (I_1 \land \dots \land I_n)$.
\end{itemize}

If we now compare algorithm $\mathcal{SE}'$ and the recursive presentation of Algorithm \ref{main_alg}
we see that they are the same with the exception that $\mathcal{SE}'$ contains redundant sub-formulas.
More critically, since we know that Algorithm \ref{main_alg} yields an interpolant of size which is worst-case quadratic in the size of the proof, we know that the same holds for $\mathcal{SE}'$ and therefore for $\mathcal{SE}$, i.e.\ for the interpolation algorithm of \cite{DBLP:conf/cade/KovacsV09}. This represents the most important downside of the approach of \cite{DBLP:conf/cade/KovacsV09} and makes it inferior to Algorithm \ref{linear_alg}.\\

Finally, interpolation from first-order refutations 
is also studied in~\cite{Bonacina15} and \cite{KV:LPAR:2017} where
the authors present 
methods for computing interpolants from arbitrary proofs in first-order
logic but either without equality or under the assumption that
colored function symbols are only constants.
While our proof splits are restricted
to local proofs, in our approach we handle first-order theories with 
equality in full generality. 


%% file: pics/digest_counterexample.tex
\begin{tikzpicture}[yscale=0.5]


        \node (bot)     at (0,0) {$\bot$};

        \node (r4)[red]      at ($(bot)+(0,1)$) {$R_4$};
        \draw [thick] ($(r4)+(-0.25,-0.5)$) -- ($(r4)+(0.25,-0.5)$);

        \node (r3)[red]      at ($(r4)+(-1.2,1)$) {$R_3$};
        \node (g6)      at ($(r4)+(0.8,1)$) {$G_6$};
        \draw [thick] ($(r3)+(-0.25,-0.5)$) -- ($(g6)+(0.25,-0.5)$);    

        \node (g3)      at ($(g6)+(-1,1)$) {$G_3$};
        \node (b1)[blue]      at ($(g6)+(0,1)$) {$B_1$};
        \node (g2)      at ($(g6)+(0.7,1)$) {$G_2$};
        \draw [thick] ($(g3)+(-0.25,-0.5)$) -- ($(g2)+(0.25,-0.5)$);          

        \node (r1)[red]      at ($(g3)+(-0.4,1)$) {$R_1$};
        \node (g1)      at ($(g3)+(0.4,1)$) {$G_1$};
        \draw [thick] ($(r1)+(-0.25,-0.5)$) -- ($(g1)+(0.25,-0.5)$);    
\end{tikzpicture}

%% file: experiments.tex

\newpage

\section{Experimental Results}
\label{sec:experiments}

We implemented \linearAlg{} (Algorithm \ref{linear_alg}, Sect.~\ref{section:local_proof_interpolation})
in automated theorem prover \Vampire{} \cite{KovacsVoronkov:CAV:Vampire:2013}
and combined it with the two approaches for obtaining a local splitting function:
the \splitByHeur{} (Algorithm \ref{alg:top-down-weighted-sum-heuristic})
and the \splitOptimally{} (Algorithm~\ref{alg:weighted-sum-optimal}). 
We will from now on refer to the combinations as \texttt{LinHeu} and \texttt{LinOpt}, respectively.
The aim of the experiment
is to compare the performance of the new algorithms to algorithm from \cite{DBLP:conf/cade/KovacsV09}
combined with its optimising improvement from \cite{HoderKV12}, 
which was already implemented in a previous version of \Vampire{}.
We will from now on refer to this latter combination as \texttt{SEOpt}.%
\footnote{
The executables and connecting scripts used in the experiment 
are available at \url{http://forsyte.at/static/people/suda/vampire_new_interpolation.zip}.}

To compensate for the lack of a representative set of benchmarks explicitly focusing on first-order interpolation,
we made use of the first-order problems from the TPTP library \cite{tptp} (version 6.4.0).
We clausified each problem using \Vampire{} and split the obtained set of clauses into halves,
treating the first half as $\mathcal{C}_A$ and the the second as $\mathcal{C}_{\neg B}$. 
We attempted to refute each of the obtained problems using \Vampire{}
(which was instructed to generate only local proofs as described in \cite{DBLP:conf/cade/KovacsV09})
and followed up by one of \texttt{LinHeu}, \texttt{LinOpt}, or \texttt{SEOpt} to compute an interpolant.
We imposed a \SI{60}{\second} time limit on the proof search in \Vampire{} and a total limit of \SI{100}{\second}
on each whole run. We ran the experiment on the StarExec compute cluster~\cite{starexec}.

In total, we obtained \num{7442} local refutations. Out of these 
\texttt{SEOpt} failed to construct an interpolant in \num{723} cases.
In contrast, \texttt{LinOpt} failed to construct an interpolant in only \num{16} cases
and \texttt{LinHeu} always constructed an interpolant within the time limit.
Furthermore, there were \num{353} cases in which \texttt{SEOpt} returned only an approximate
result and \num{108} cases where optimisation failed and the 
unoptimized version of \cite{DBLP:conf/cade/KovacsV09}
was used as a fallback instead. The observed higher computational demands of \texttt{SEOpt}
can be mostly ascribed to the reliance on a different pseudo-boolean solver 
and different connecting technology.\footnote{
\texttt{SEOpt} uses the SMT solver Yices \cite{DBLP:conf/cav/Dutertre14} (version 1.0) and communicates via a file,
while \texttt{LinOpt} one relies on Z3 \cite{Z3} (version 4.5) and its API.}
These differences 
unfortunately exclude the possibility of a meaningful 
comparison of more detailed timing results. 
However, we would like to point out that the optimisation problem \texttt{SEOpt} constructs
is arguably much more complex than the one stemming from \splitOptimally{} employed by \texttt{LinOpt}.

\begin{figure}[t]
\centering
\includegraphics[scale=0.43]{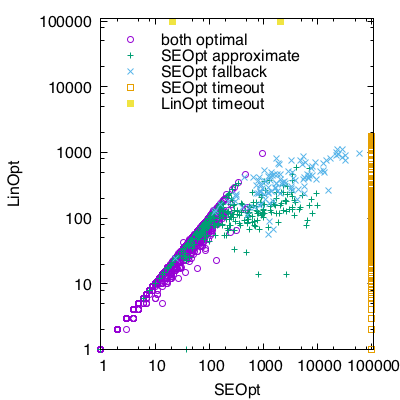}\includegraphics[scale=0.43]{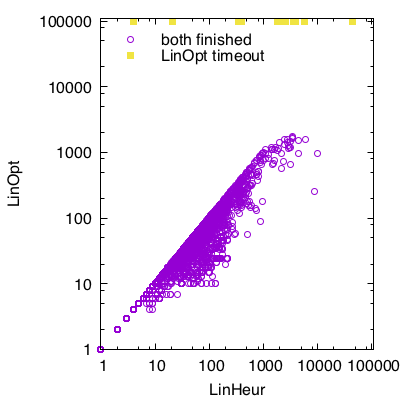}
\caption{Size comparison of interpolant produced by \texttt{SEOpt} and \texttt{LinOpt} (left), and \texttt{LinHeu} and \texttt{LinOpt} (right). Each point corresponds to a single refutation and its position to the sizes of the respective interpolants.
}
\label{fig:main_comparison}
\end{figure}

Fig.~\ref{fig:main_comparison} (left) contains a scatter plot comparison of the sizes of obtained interpolants
for \texttt{LinOpt} and \texttt{SEOpt}.
An artificial large value was substituted whenever a particular algorithm failed to provide an interpolant.
This is reflected by the data points on the right and the upper border, respectively.
The plot further separates the points to categories based on the optimality guarantee provided by \texttt{SEOpt}.
We can see that \texttt{LinOpt} yields consistently better results.
Moreover, the improvement
tends to get more pronounced with the growing size of the instances. Finally,
even when just focusing on instances where \texttt{SEOpt} finished optimising, there are numerous cases
where the interpolant from \texttt{LinOpt} is several times smaller than that of \texttt{SEOpt}.
This is because \texttt{SEOpt} cannot avoid repeating certain formulas from the refutation many times in the interpolant
and corresponds to the worst case quadratic complexity discussed in Sect.~\ref{sec:related}.\footnote{
An interesting side-effect is an ability of \texttt{SEOpt} to assign two different colors
to a formula when considered from the perspective of two different sub-derivations.
In rare cases, such formula does not need to appear at all,
and the final interpolant may be smaller than what is achievable by \texttt{LinOpt}.
An instance of this phenomenon occurred in our experiment on benchmark \texttt{SYN577-1},
which appears in Fig.~\ref{fig:main_comparison} (left) slightly above the diagonal.}


Fig.~\ref{fig:main_comparison} (right) correspondingly compares \texttt{LinOpt} with \texttt{LinHeu}.
Although the plot highlights many examples where \texttt{LinHeu} yields a larger interpolant than \texttt{LinOpt},
an optimal interpolant is actually discovered by \texttt{LinHeu} in \SI{79}{\percent} of the cases
and its interpolants are only \SI{11.6}{\percent} larger on average. Moreover, 
on the 7429 refutations on which both algorithms finished in time,
the accumulated time spent on interpolant extraction by \texttt{LinHeu}
was only \SI{8.17}{\second} compared to a total of \SI{1901.03}{\second} spent by \texttt{LinOpt}.
This shows that \texttt{LinHeu} presents a viable alternative to \texttt{LinOpt}
when trading the quality of interpolant for computational time becomes desirable.

Given the encouraging results we intend to officially replace \texttt{SEOpt} by \texttt{LinOpt} and \texttt{LinHeu} in \Vampire{}
and make it available with the next release of the prover.








%% file: conclusion.tex
\section{Conclusion}
\label{sec:conclusion}

We presented a new technique for constructing interpolants from first-order local refutations.
The technique is based on an idea of proof splitting and on a novel non-inductive construction
which arguably gives more insight than previous work and yields interpolants of linear size.
This leads to a new interpolation algorithm which we implemented in the automated theorem prover \Vampire{}.
Finally, we confirmed in an extensive experiment that the algorithm also improves over the state-of-the-art in practice.
